\documentclass[5p,times,twocolumn]{elsarticle}

\journal{}

\begin{document}
\begin{frontmatter}


\title{Analytical Observations on Knapsack Cipher 0/255}
\author{Ashish Jain, Narendra S. Chaudhari}
\address{Department of Computer Science and Engineering, Indian Institute of Technology Indore, India}
\ead{ashishjn.research@gmail.com, nsc@iiti.ac.in}

\begin{abstract}
We observed few important facts that concerns with the new proposal of knapsack cipher 0/255, recently published by Pham \cite{pham2011improvement}. The author claimed that the time complexity for solving new improved trapdoor knapsack is O(${256}^{N}$). In this paper, we show that the knapsack cipher 0/255 can be solved in the same time that is required  for solving the basic knapsack-cipher proposed by Merkle and Hellman \cite{merkle1978hiding}. In other words we claim that the improved version proposed by Pham \cite{pham2011improvement} is technically same as the basic Merkle and Hellman Knapsack-based cryptosystem.
\end{abstract}


\end{frontmatter}

\section{Introduction}
\label{intro}
The trapdoor knapsack used for hiding information and signature is a knapsack-based cryptosystem, first proposed by Merkle and Hellman \cite{merkle1978hiding} in 1978. This public key encryption proposal has been thoroughly investigated owing to a high computational efficiency (at that time). The motivation of its design is converting $superincreasing$ knapsack sequence into a computationally hard sequence. Though, the basic version was broken by Shamir in 1984 \cite{shamir1984polynomial}. Despite of failure of previous all knapsack Public Key Cryptosystems (PKC), two new knapsack PKC proposed by Wang et al \cite{wang2007knapsack} and Murakami et al \cite{murakami2008new} in 2007 and 2008, respectively. However, very recently both has been broken by peng at al \cite{peng2013analysis} by mounting lattice-based attack.\\
\indent Shor \cite{shor1997polynomial} showed that, the security of most PKC proposed so far depends on the difficulty of integer factorization problem or discrete logarithm problem. However, these problems can be easily solved using quantum computers. Knapsack problem is one that cannot be easily solved using quantum computers.\\ 
\indent Now a days there is requirement of light weight, high-speed and highly secure cryptography algorithms for electronic commerce \cite{hamilton1997commerce}. So, recently an improvement to the basic knapsack cryptosystem is proposed in the captioned paper ``The improvement of the knapsack cipher" by Pham\cite{pham2011improvement}. However, during the critical analysis of \cite{pham2011improvement}, it was observed that even though the author claimed the time complexity to be O($256^{N}$) for solving trapdoor knapsack 0/255, we show that the exhaustive search will take the same time that is required by the basic trapdoor knapsack proposed by Merkle and Hellman \cite{merkle1978hiding}. To the best of our knowledge, this the first paper, that presenting analytical observation on knapsack cipher 0/255.\\ 
\indent The remainder of the paper is organized as follows: In the Sect.~\ref{sec:2} knapsack cipher 0/255 is described, comments on the knapsack cipher 0/255 is given in Sect.~\ref{sec:3} followed by conclusion in Sec.~\ref{sec:4}.

\section{Description of the Knapsack cipher 0/255}
\label{sec:2}

In any PKC system, public key is publicized by the designer (e.g. by Alice), so that using public key sender (e.g. Bob) encrypts the plaintext and sends it over a  (insecure) communication  channel. Upon receiving the encrypted message (ciphertext $b$), the receiver (Alice) decrypts the ciphertext $b$ using her own private key ( A key that is used by the designer to generate the public key). In the case of Merkle-Hellman knapsack-based PKC, the trapdoor knapsack $A=a_1, a_2,..., a_n$ (set of natural numbers) is publicized as the public key (A typical value of $n$ is 100 and a typical size of each $a_i$ is 200 bits). Let, the sender have a message of length $n$ as a bit string or simply $X = x_1, x_2,..., x_n$ \{$x_{i} \in{0,1}$\}. The sender first compute the sum $b$ and then sends it via the public channel. \\

where: $b =\sum_{i=1}^n a_{i}*x_{i}$\\

Both the receiver and the potential eavesdropper knows the public encryption vector $A$ and the ciphertext $b$. Their task is to find which subsets of the $a_{i}$  sums up to $b$. This is an instance of the knapsack problem, which is known to be nondeterministic polynomial time complete (NP-Complete). This problem is difficult for the eavesdropper but easy for the receiver because she have the private key ($A^{'}, w^{'}$ and mod $m$) i.e. easy knapsack, inverse of multiplier and modular $m$ respectively.  

To achieve a compromise between speed and security, the vector $A$ of size $n$ can be cut to a small size by a factor of $1/f$ (i.e. $N=n*(1/f)$) without changing the size of $X$ ( i.e. n). Since the size of $A$ is reduced but the size of $X$ remains unchanged, we must allow each $x_{i}$ to take on values from the set \{0,1,2,..., $2^{f}$-1\}. These modifications are possible if the designer performs the following steps.\\

\begin{enumerate}
  	\item The designer chooses a $superincreasing$ vector $A^{'}= (a^{'}_{1}, a^{'}_{2},..., a^{'}_{N})$. Let, $N=n*(1/f)$
	\item Select a modular $m$ $>\sum_{i=1}^{N} a^{'}_i$.	
	\item Choose a multiplier $w$ in between 1 and $m$-1 so that ($w$, $m$) must be co-prime.
	\item Generate a vector $A$ of size $N$ as $a_1, a_2,..., a_N$ (here, $a_i=w*a^{'}_i$ mod  $m$).
	\item Publicize the vector $A$.	
\end{enumerate}

It is clear from the above steps that the result of modification is a reduction in the volume of transmitted data without changing the size of vector $X$. As a result, the time required for transmitting data is reduced by a factor of 1/f. However, an important fact is that the complexity of solving trapdoor knapsack $A$ is remains unchanged by allowing the above modification.

If the size of vector $X$ is 96 and f=8 then N=96/8=12, that allows each $x_{i}$ to take on values from the set \{0,1,2,..., 255\}. Actually, the knapsack cipher 0/255 is ($A^{'}$) and the trapdoor knapsack is ($A$) in the captioned paper ``The improvement of the knapsack cipher"  proposed by Pham \cite{pham2011improvement}.

\section{Comments on the Knapsack Cipher 0/255}
\label{sec:3}
It is noteworthy that the knapsack problem is more general. In fact, the knapsack-based cryptosystem is a specific instance of the knapsack problem that is called integer partitioning problem. \\
\indent During critical analysis of Merkle and Hellman paper\cite{merkle1978hiding}, we observed an important fact in section V (Compressing the Public File).
$n$=100 is the bottom end of the usable range for secure system. But, to maintain a balance between speed and security, the vector $X$ must be 100 bits long while $n$ can be reduced to say 20 ( $N$=20). As a result, the transmitted data is reduced by a factor of five (100/20). That is possible by allowing each element $x_{i}$ to take on values in the set \{0, 1, 2,...,31\} instead of \{0,1\}. However, the original equation 1 must be modified to equation 2.
\begin{equation}
a^{'}_{i} > \sum_{j=1}^{i-1} a^{'}_j
\end{equation}
\begin{equation}
a^{'}_{i} > 31*\sum_{j=1}^{i-1} a^{'}_j
\end{equation}
$Example:$\\ 
Transmitting 20 Kbits on a low-speed 300 bit/sec takes more than a minute. But if the transmitted data is reduced by a factor of 8 to about 2.5 Kbits. Then, the transmission process will  takes less than 8 seconds. This is accomplished by reducing the number of $a_i$ to 12 elements \footnote{If the reductions in number of $a_i$ is represented by $N$ and the size of vector $X$ is 96. then $N$=96/8=12.}. Since the size of vector $X$ is 96, then for each element $a_i$, we must reserve 8 bits in vector $X$ i.e. each $x_i$ to take any values in 0 to 255 ($2^{8}-1$) .

\subsection{Time Complexity for solving trapdoor knapsack in the worst case}
Let $n$ is the length of vector $X$.\\

Case-I: If the length of publicized vector $A$ keep same as the length of the vector $X$ \{$x_{i} \in{0,1}$\}, then trapdoor knapsack $A$ can be solved in time O($2^{n}$).\\
e.g. Let, length($X$)=length($A$)=96 and $x_{i} \in \{0,1\}$. Then, the time required for searching solution exhaustively=O($2^{96}$).\\  

Case-II: If the length of publicized vector $A$ is reduced as (1/f)*length($X$), then $x_{i} \in \{0,1,2,..., 2^{f}-1\}$. As a result, the trapdoor knapsack $A$ can be solved in time O(($2^{f})^{N})$, here $N=n*1/f$.\\
e.g. Let, length($X$)=96 and $x_{i} \in \{0, 1,..., 255\}$, then $N$=12. However, the time taken for searching solution exhaustively=O(($2^{8})^{12})$.\\\\   
From Case-I and Case-II, it is clear that\\ 
\indent O($2^{96}$)=O(($2^{8})^{12})$\\
\indent\hspace{10 mm}=O($256^{12}$)\\
In general:
O($2^{n}$) = O($256^{N}$).\\ where $N$=$n$/8,  i.e. the reduction in number of elements by 1/8 allowing the hike in speed by 8.

\section{Conclusion}
\label{sec:4}
The author of \cite{pham2011improvement} has defined a ``super-increasing vector level 2" as $V^{'}=(v_{1}, v_{1},... ,v_{n})$. If we keep the length of publicized vector $V$ and the vector $X$ is the same, but, since $x_{i} \in \{0,1,... ,255\}$, such a knapsack cryptosystem is practically not possible. In the knapsack cipher 0/255,  the author has not taken attention on the speed of cipher, that is very important factor for secure communication. However, if the size of vector $V$ is reduced to a factor(1/f), then it results in a hike for speed by a factor(f). An important fact is that even the transmission speed will be improved, the efficiency of solving trapdoor knapsack remains the same. We would like to add some more facts is that the first serious attack on the basic version of Merkle and Hellman cryptosystem was mounted by shamir in 1984 \cite{shamir1984polynomial} by exploiting the special structure of the sequence of knapsack. The basic tool for analysis was sawtooth curves (function of $Va_i(mod \hspace{1 mm} m)$). In which, accumulation points of the minima of $l$ sawtooth curves is found by dividing both coordinates of the curve by modular $m$. In this way, we get the sawtooth curve of the function of $Va_{i}(mod \hspace{1 mm}1)$. Since, the function is now independent to $m$ and 0 $ \leq 
V<1$, the attack is applicable to any knapsack cipher 0/$2^f$.

\bibliographystyle{elsarticle-num6}
\bibliography{myref}
\end{document}